\documentstyle[11pt,aaspp4]{article}

\slugcomment{To appear in ApJ, part 1}

\lefthead{Gioia et al.}
\righthead{The lensing cluster MS0440+0204}
\epsfverbosetrue
\def\deg{\hbox{$^\circ$}}
\def\sun{\hbox{$\odot$~}}
\def\lefevre{Le\thinspace F\`evre~}
\def\etal{{\it et al.} }
\def\ms04{MS0440$+$0204}
\def\hawaii{Hawai$'$i~}

\begin{document}

\title{The Lensing Cluster MS0440+0204 Seen by HST, ROSAT and ASCA:
I. Cluster Properties\altaffilmark{1}}

\author{I. M. Gioia\altaffilmark{2,3,4}, E. J. Shaya\altaffilmark{5},
O. \lefevre\altaffilmark{6},  E.E. Falco\altaffilmark{7},
G.A. Luppino\altaffilmark{2,4} and
F. Hammer\altaffilmark{6}}

\altaffiltext{1}{Based on observations made with the NASA/ESA Hubble
Space Telescope, obtained at the Space Telescope Science Institute,
which is operated by AURA under NASA contract NAS5-26555, as well as
on observations made with the Multiple Mirror Telescope
Observatory, which is operated jointly by the University of Arizona
and the Smithsonian Institution}

\altaffiltext{2}{Institute for Astronomy, 2680 Woodlawn Drive,
Honolulu, HI 96822}

\altaffiltext{3}{Istituto di Radioastronomia del CNR,
Via Gobetti 101, I-40129, Bologna, ITALY}

\altaffiltext{4}{Visiting Astronomer at CFHT,
operated by the National Research Council of Canada, le Centre
National de la Recherche Scientifique de France and the University
of \hawaii, and at the W. M. Keck Observatory, jointly
operated by the  California Institute of Technology and the
University of California}

\altaffiltext{5}{Physics Department, University of Maryland, College
Park, MD 20742}

\altaffiltext{6}{DAEC, Observatoire de Paris Meudon, 92195 Meudon
Principal Cedex, FRANCE}

\altaffiltext{7}{Center for Astrophysics, 60 Garden Street, Cambridge,
MA 02138}

\begin{abstract}

We present an analysis of the properties of the lensing
cluster \ms04 at z$=$0.1965.  \ms04 has been observed with a
variety of telescopes at diverse wavelengths: from
the ground with {\it CFHT, MMT} and {\it KECK} and from Earth
orbit with {\it HST, ROSAT} and {\it ASCA}. Mass determinations
are separately obtained from galaxy virial motions and X-ray profile
fitting. A simple $\beta$-model fit to the X-ray data yields a mass of
(1.3$\pm0.2) \times10^{14} M_{\sun}$ within 583 kpc of the cluster
center, but more general models fit all of our data better and
allow a wider range of masses that are consistent with the lensing
data.  In addition, the X-ray data yield a mass distribution
profile that is well described by a $\beta$ model with a core radius of
26.7 kpc.  The velocity dispersion of galaxies yields
a  mass of 4.8$^{+1.5}_{-0.94} \times10^{14} M_{\sun}$
within 900 kpc.  In the inner 24\farcs5 there are
24 arcs that appear to be  strong gravitationally lensed images of
background sources.  Models of the cluster mass distribution and its
lensing properties reveal 5 background sources at various
redshifts  each forming 2 or more arcs. We do not have a redshift
for any arc with multiple images, therefore we can only place upper
and lower limits to the mass of the cluster from gravitational
lensing.  At 100 kpc, the lower limit mass from lensing is about a
factor of 2 greater than the X-ray determined mass.  The rate of
increase in the projected mass at this radius also is greater for the
lens model than the X-ray determination. To reconcile the mass
estimates from the X-rays and the lensing and to try to understand
the steep slope of the gravitational lens mass, we tentatively
explore a model with a supercluster surrounding the cluster
and with a mass profile that increases more rapidly than a
$\beta$ model at large radii.

\end{abstract}


\section{Introduction}

As the largest gravitationally bound structures known, clusters
can set clear constraints on the formation of structure and on the
composition of the universe. Three independent techniques
have been used to determine the mass distribution in galaxy clusters.
The oldest and most conventional approach is based on galaxy velocity
dispersion and application of the virial theorem.  Here it is assumed
that galaxy orbits are isotropic and that light traces the dark matter.
A second method  derives gravitational mass profiles from X-ray
observations under the assumptions of thermal hydrostatic equilibrium
and spherical symmetry. In this technique only two observables
must be known to reconstruct the distribution of mass, the electron
density of the cluster gas and its temperature, clearly an advantage
with respect to the often limited optical data available to define
the cluster structure. The X-ray derived mass, however, may not
necessarily be representative of the true cluster mass if merger
compression or shocks are present. Over the last ten years it has been
possible to determine cluster masses using the effect of gravitational
lensing, both in its strong  and weak manifestations (see the excellent
review of Fort and Mellier, 1994, for a list of clusters that
has been analyzed.).
The lensing method has the advantage  that the mass measurement is
independent of the thermodynamical state of the gravitating matter.
The weak distortions (Tyson \etal 1990; Kaiser and Squires,
1993) are particularly suited to map cluster mass at large radii.
Strong lensing is restricted to the cluster core.

The use of the three techniques in conjuction allows an examination
of the uncertainties of each method and provides a unique possibility
to study the dynamical and physical state of the gas and dark matter
in clusters. It is worth noting that even if theoretically
one should obtain the same masses if clusters are dynamically
relaxed, in practice comparisons of masses derived from dynamical
analyses and  gravitational lensing have shown a significant
discrepancy in the mass estimates (M$_{lens}/M_{dyn} \sim$ from 5 to 2
going from inner 250 -- 300 kpc of the cluster center up to 1 Mpc;
see Wu and Fang, 1996, 1997 and references therein), but there are
exceptions as mentioned below.

Several possibilities have been suggested to resolve this discrepancy.
Among others: inadequacy of the isothermal, hydrostatic equilibrium
models of the X-ray analyses that cause systematic underestimate of
the cluster mass (Wu and  Fang 1996); projection effect of an asymmetrical
matter distribution (Miralda-Escud\'e and Babul, 1995); presence of
substructures in X-ray clusters and cluster mergers (Henry and Briel
1995, 1996; Markevitch 1996) which would help to explain the discrepancy
in A2218 (Kenib \etal 1995; Squires \etal 1996a) and in A2219 (Smail
\etal 1995a); inhomogeneous intracluster medium (Miralda-Escud\'e and
Babul, 1995); the possibility of X-ray lensing  (in A2218, Markevitch
1997); existence of nonthermal cosmic ray pressure which could support
the intracluster ionised gas (Ensslin \etal 1997); offsets between
X-ray and lensing centres and overestimate of the core radii of the
dominant mass clumps in non-cooling flow systems (Allen, 1997).
However, recent analyses with better data show a more complex picture
that will lead to a better understanding of the physical processes going
on in clusters.

Marginal agreement of the mass determinations from X-ray and lensing
analyses is found in A2163 (Miralda-Escud\'e and Babul 1995), and
confirmed  by Squires \etal 1997, using weak lensing  and
new ROSAT HRI and PSPC data. A similar analysis by Smail \etal 1995b,
for the Medium Survey clusters MS1455$+$22 and MS0016$+$16 found
agreement between the lensing and X-ray masses. A consistent picture
is constructed for A2390 by Pierre \etal 1996, and  Squires
\etal 1996b. Allen \etal 1996, present a
multi-phase X-ray analysis of PKS0745$-$191, a regular and relaxed
cluster with a massive cooling flow. The excellent agreement of the
mass distributions lead them to conclude
that the X-ray gas in PKS0745$-$191 is in hydrostatic equilibrium, and
that non-thermal pressures components are not required by the data,
differently from the cases of A1689 or A2218 (Miralda-Escud\'e
and Babul, 1995; Loeb and Mao 1994). In a recent analysis of 13 clusters
observed with ASCA and ROSAT, Allen points out that the X-ray and strong
gravitational lensing mass measurements show excellent agreement  for
the cooling-flow clusters in his sample, while for the non-cooling
flow clusters, the masses determined from the strong lensing data
exceed the X-ray values by factors of 2$-$4. Allen suggests that
these discrepancies can be reconciled if one takes into account
that the dynamical activity observed in non-cooling flow clusters
has caused the X-ray analyses to overestimate the core radii of the
dominant mass clumps. Other factors as substructure and line-of-sight
alignments of material towards the cluster cores may also contribute
to the discrepancies. A quite different and interesting
approach is taken by Smail \etal (1997) to analyze a small sample of
12 clusters observed by the Hubble Space Telescope (HST). From the
comparison of the mean gravitational shear strength with the
cluster X-ray luminosities they develop a model used to predict the
relationship expected from properties of local clusters.  In this way
they can distinguish between models for the evolution of the cluster
properties. It is an innovative and promising study to measure
cluster evolution once an expanded and better defined sample of
clusters are examined.

In this paper, we present a study of the gas and mass distributions of
the cluster of galaxies \ms04. Originally discovered through its X-ray
emission in the Extended Medium Sensitivity Survey (EMSS; Gioia \etal
1990), \ms04 was part of a Mauna Kea based observational program to search
for arcs and arclets in a complete sample of X-ray luminous medium-distant
(0.15$\leq$z$\leq$0.83) clusters of galaxies.  At a redshift of
z$=$0.1965, \ms04 has the most striking example of an arc system in a
compact, centrally condensed cluster. Ground based CCD observations of
\ms04 show at least 15 blue segments of circular structures surrounding
a multiple nucleus cD galaxy (Luppino \etal 1993). The arcs are
unresolved even in superb observing conditions (seeing $\sim$0.5$''$) at
Mauna Kea Observatory.

We have extended the study of \ms04 with the refurbished HST and with
X-ray satellite observations. Deep images, acquired with the WFPC2
camera aboard HST, reveal detailed structures in both the previously
known arcs and the newly discovered arcs. Constraints on the mass of
the cluster are derived from detailed modeling of these arcs. Additional
mass estimates are obtained from X-ray observations of \ms04 with
ROSAT/HRI and ASCA. We also acquired spectra for 40 cluster members from
the ground and, therefore, were able to estimate the velocity
dispersion of the cluster, which yields an independent dynamical estimate
of the cluster mass.  We present both the optical and X-ray data that we
have acquired and make comparisons between the diverse and complementary
mass distribution estimates for \ms04. To reconcile the discrepancy of
mass estimates from the dynamical and lensing analyses, we present a very
simple model with a mass profile that increases more rapidly than a
$\beta$ model at large radii. The model explored in this paper
is one with two isothermal spheres. We remind here that more general
models fit all of our data better and allow a wider range of masses
that are consistent with the lensing data (Shaya \etal 1998).
We emphasize that the two isothermal spheres model is
indeed a speculation
but it is presented as one of the possible solutions to remove the
discrepancy. Subsequent papers will give a more detailed analysis of
the cluster properties and mass models that we are studying based on
our multi-frequency dataset. Throughout this paper, we assume H$_{0}=50$
h$_{50}$ km s$^{-1}$ Mpc$^{-1}$, the density parameter $\Omega = 1$,
and the cosmological constant $\Lambda=0$, unless otherwise stated.
At the redshift of the cluster, the luminosity distance is 1231
h$_{50}^{-1}$ Mpc, the angular size distance is 860 h$_{50}^{-1}$ Mpc,
and the scale is 4.17 h$_{50}^{-1}$ kpc per arcsec.

\section{HST Imaging}

\subsection{Observations and photometry}

We acquired 10 exposures on consecutive orbits with the WFPC2 and the
F702W filter in October 1994, for a total integration time of 22,200~s.
The core of the cluster fits conveniently inside the $1\farcm3 \times
1\farcm3$ field of view of the Wide Field Camera 3 (WFC3), the best
performing chip. The pixel size in this camera is  99.6
milliarcsec (Holtzman \etal 1995). Each exposure
was offset by an integer number of pixels in both axes to aid in the
correction for cosmic rays, dead pixels, and hot pixels. The standard
STScI processed frames were registered and co-added using IRAF/STSDAS
routines. The final 4-chip mosaic frame
is shown in Figure 1 (Plate). The diffuse light,
mostly from the envelope of the cD galaxy and partially from other
galaxies, is detectable everywhere in the lensing region of this
cluster. In Figure 2 (Plate), the diffuse light from the
galaxies is diminished to bring out sharp structures by subtracting an
image composed of the median value in a 19 by 19 pixel box around each
pixel.

The photometry was performed using the faint galaxy photometry software
described in Le F\`evre \etal (1986), and Lilly \etal (1996). The
photometric calibration was based on the HST calibration coefficients
(Holtzman \etal 1995) and confirmed by our ground based imaging (Luppino
\etal 1993) within 0.1 magnitude.  Manual intervention was required to
include the arcs with the largest axis ratio, which were not identified
by the software. The histogram of number counts with Johnson R band
magnitude is presented in Figure 3. The number counts decline for
$R\geq26$ indicating that the counts are incomplete at fainter
magnitudes. A total of 901 objects with peak intensity above
$\mu_{R}=$25.5 mag arcsec$^{-2}$ ($3\sigma$ over the sky background)
have been identified in the 4.71 arcmin$^2$ HST field usable for
photometry. Table 1 gives the R surface brightness in mag
arcsec$^{-2}$ for the objects with available spectroscopy (see Section
3. for a description and  Figure 4).
The peak of the light distribution has been determined after removal
of the bright star next to the core. After a gaussian filter
with $\sigma=20$ pixels was applied to the image, the peak of the
light distribution is measured to be 1\farcs6 from  galaxy A.
Ellipse fitting of the light envelope of the central core  for the
$\mu_R=24.25$ mag arcsec$^{-2}$ isophote indicates a major axis of
25\arcsec, an ellipticity of $\epsilon=0.17$, with
$\epsilon = (a^2-b^2)/(a^2+b^2)$, and a position angle of
78$\deg$. The center of this isophote fitting, indicated by a cross in
Figure 2, is at $\alpha$~=~04$^{h}$43$^{m}$09\fs71 and
$\delta = +02\deg10\arcmin18\farcs66$ (J2000), just 1\farcs2
North and 1\farcs1 West from galaxy A.

\subsection{Arcs and  Arclets}

The arc system is remarkable for the symmetry of the distortion
pattern, and the large number of very elongated arcs. The most
spectacular arc system is formed by arcs A2 and A3 (see
Figure 2) which are resolved into bright knots by the HST.
The very high axial ratios of these arcs indicate that they are quite near
the critical radius.  Arc A1, which was the most spectacular looking arc
in the ground based image, appears to be a highly distorted image of a
galaxy but not necessarily a strongly lensed object and we have found no
other counter images of it.  Arcs A5 and A6 both have multiple knots.
Each of these knots, when reconstructed at the source plane, merges
together with its counter image.  The ability to bring these two
intricate arcs together in a consistent way is a key requirement for a
successful reconstruction for this cluster. Arcs A8 and A9 are very
close to each other and are at nearly the same distance from the
center of the light distribution.  It is most probable that they
emanate from a single object near the critical radius.
The critical radius in the image plane grows with the redshift of the
source, therefore the source of arcs A8 and A9 must be at a lower
redshift than the source for arcs A2 and A3.

\subsection{Radial Arcs}

Two radial structures are observed near the center of the cluster.
To identify the geometry of each feature more clearly  and measure
its magnitude, we have processed the central region of the cluster
in the following manner. The two galaxies at the
northern edge of the radial arc A17 have been modeled with $r^{1/4}$
profiles, and then subtracted from the image. A gaussian filter with
$\sigma=2$ pixels was then  applied to the resulting image, and the
result was subtracted from the non-filtered image.
One arc is the slightly curved feature (A17 in Figure 2)
6\farcs5 to the North of galaxy A. It subtends 2\farcs5, and has a magnitude
$R=25.6\pm0.3$. Another radial feature is observed 4\arcsec\ North of
galaxy B (A16 in Figure 2). It subtends 2\arcsec\, with
$R=27.2\pm0.6$ and is thus a more marginal feature. Both radial
arcs have been used in the reconstruction to background source
galaxies of this cluster.

\section{Spectroscopic data}

Spectroscopy was obtained in the field of \ms04 in October 1993 with the
Canada-France-Hawaii Telescope (CFHT) equipped  with the Multi-Object
Spectrograph (\lefevre \etal 1994). The O300 grism was used to provide a
wavelength coverage from 4500 \AA\ to 9000 \AA and a pixel size
of 5.5 \AA/pixel. A slit width of 1\farcs5
was used, providing a  spectral resolution of 17 \AA. Each 30 minute
exposure was offset to allow removal of cosmic rays and bad pixels. The
total exposure times for each object ranged from 2 to 3.5 hrs depending
on which masks the object was seen through and on weather conditions
(mean seeing $\sim$ 1\farcs5). The data were reduced using the MULTIRED
package developed by \lefevre \etal  (1995). Fourteen galaxies, out of
15 objects observed, are cluster members. One galaxy is the arc-like
structure to the East of the cluster core (A1 in Figure 2).

\ms04 was also observed in November 1993, at the Multiple Mirror
Telescope (MMT) with the Red Channel spectrograph and the 300 gr/mm
grating.
Exposures between 1 and 1.5 hrs were made using 7 aperture plates with
between 7 and 10 slits each. The pixel size was 3.21 \AA/pixel, the
spectral resolution was 10 \AA, and the seeing was
between 1\arcsec\ and 1\farcs5. Standard data reduction and velocity
measurements with the XCSAO package under IRAF confirmed 37 cluster
members. Nine cluster galaxies (labelled as G6, G10, G13, G18, G26,
G28, G30, G33 and G35 in Figure 4) are in common
with the CFHT observations, thus allowing an estimate of external
errors. The average velocity difference from the common data
is $<v_{\rm(MMT)}-v_{\rm (CFHT)}>=$227 km s$^{-1}$  with an
$rms$ dispersion of 327 km s$^{-1}$.

We attempted to obtain long-slit spectroscopy of the brightest arcs in
MS0440$+$0204 in January and October 1995 with the Low Resolution
Imaging  Spectrograph (LRIS) at the W. M. Keck 10m telescope. Poor
weather and problems with the instrumentation prevented us from
obtaining additional useful spectra of arcs. However,  part of the
data from both runs could be used to obtain spectra for 10 objects.
These objects  happened to fall in the $1\arcsec\times3\arcmin$\ or
$1\farcs5\times3\arcmin$\ slits that were positioned on the arcs.
The 300 gr/mm grating was used which gives a pixel size of
2.5 \AA/pixel. Four objects are cluster members (1 in common with MMT
data; G8 in Figure 4), one object at $\sim$ 1\farcm6 to
the Southwest of the cluster is a QSO, another is an M star, and the
remaining objects are galaxies in the background of the
cluster. Their redshifts range from 0.387 to 0.777, but none of these
galaxies lies within the effective strong-lensing area.

All the objects with spectroscopy are listed in Table 1 and marked
in Figure 4. For each object the measured velocity
(barycentric) plus its 1$\sigma$ error, and redshift are given.
Galaxies marked as G, C, and K in Figure 4 have been
observed with the MMT, the CFHT, and the Keck telescope, respectively.
K5 denotes the QSO in the field at z=0.4719$\pm$0.0006.
Velocities are given for eight additional galaxies (marked 1 through 8)
which are not cluster members (redshifts range from 0.0766 to 0.2617).
Galaxy C4 is outside the field of view of Figure 4.

In total, 54 objects have securely identified redshifts in the field
and 40 objects are consistent with being cluster members (marked with
an asterisk in Table 1). The velocity histogram for the 44 galaxies
with the approximate redshifts of \ms04 is shown in
Figure 5.
There is a low-velocity extension in the histogram at 55,000
km s$^{-1}$ (4 galaxies) that our 3$\sigma$ clipping iterative
algorithm (following Danese \etal 1980) excludes from the computation
of the cluster velocity. From 40 accepted cluster
members, we obtain a mean velocity of $<$v$>$ = 58,909$\pm$142
km~s$^{-1}$ and a dispersion along the line of sight of
$\sigma_{los} = 872^{+124}_{-90}$ km s$^{-1}$.
Carlberg \etal 1996 found a value of 606$\pm$62 km s$^{-1}$ for
the velocity dispersion of \ms04.
They attribute their lower determination of velocity dispersion in
a number of clusters to three factors (see their paper, section 3.3).
Among them is the fact that the larger radial range covered by their
data makes their data less vulnerable to the presence of local
substructures which might affect the velocity dispersion.

Figure 5 includes a gaussian distribution centered
at the redshift of the cluster, of the appropriate width, and with
area normalized to the 40 cluster galaxies.  From these data, the
redshift of the cluster is measured to be 0.1965$\pm$0.0005.

The CFHT optical spectrum for galaxy A1 (see Figure 2
and Table 1) is shown in Figure 6. The presence of four
emission lines in the spectrum unequivocally identifies this galaxy
as a lensed background galaxy to the cluster at z=0.5317$\pm$0.0003.

\section{X-ray data}

\subsection{ASCA temperature}

\ms04 was observed by ASCA (Tanaka, Inoue \& Holt, 1994) in September
1994 for 40,000~s.  The ASCA  instrumentation consists of two
Solid-State Imaging Spectrometers (SIS) sensitive in the range 0.5$-$9
keV (140 eV resolution at E$=$6 keV), and two Gas Imaging Spectrometers
(GIS) with poorer energy resolution but with some efficiency up to 11
keV. The SIS observations were performed in 4 CCD-Bright mode for
Medium bit rate observations.
Data preparation and analysis were done using the XSELECT and
FTOOLS software packages which allow selection of valid time intervals
and removal of hot and flickering pixels. Additional analysis was
performed using the XSPEC package.  For the spectral analysis we use the
summed spectra of the two GIS and of SIS0. SIS1 spectrum was not used
because of a contamination problem in the detector.
The spatial resolution of ASCA along with the low
signal-to-noise of the data permitted only a global single-temperature
Raymond-Smith model (Raymond and Smith 1977) to be fitted to the data.
The hydrogen column density was fixed to the Galactic value along the
line of sight at the cluster position, N$_{H}=9.12\times10^{20}$
cm$^{-2}$ (Stark \etal 1992), and the heavy elements abundance was fixed
to 0.3 of the solar value. This value was chosen since the typical range
of measured abundances for clusters with  temperatures between 2 and 6
keV lie between 0.3 and 0.4 solar (Ohashi 1996). The region containing
the QSO could not be excluded from the analysis given its proximity to
the cluster centre (1\farcm6, less than the half power diameter of
the point spread function of the ASCA XRT +GIS, Makishima \etal 1996).
A temperature of
$kT=5.5^{+0.8}_{-0.6}$ keV (90\% confidence interval) was measured for
the cluster gas. The inclusion of the QSO in the analysis would raise
the fitted cluster temperature by only 6.5\%. This variation is within the
temperature errors and it is thus negligible for our purposes.

\subsection{ROSAT imaging}

X-ray observations of \ms04 were obtained with the ROSAT High Resolution
Imager (HRI, Tr\"umper 1983) in two pointings in February 1994 and in
August 1995, for a net live time of 27,015~s. The HRI operates in the
ROSAT 0.1$-$2.4 keV energy band and provides an angular resolution of
$\sim$~4\arcsec\ (FWHM).

X-ray iso-intensity contours of \ms04 are shown in
Figure 7 overlaid on an optical CCD frame. We first
created an X-ray image with a pixel size of 1\arcsec\ and then smoothed
the image with a Gaussian of $\sigma$~=~4\arcsec. The coordinates of the
X-ray centroid are $\alpha$(J2000)~=~04$^{h}$43$^{m}$09\fs8 and
$\delta$(J2000)~=~02\deg10\arcmin19\farcs5, corresponding to a position
2\arcsec\ North of galaxy A, and within 1\farcs5 from the peak of
the light distribution. The X-ray emission has average ellipticity
$\epsilon=0.15\pm$0.02. The position angle of the major axis with
respect to the North axis (counterclockwise positive) is PA$=$120\deg\
for the innermost isophotes,  and PA$=80\deg$ ~for the outer isophotes.
The center of  iso-intensity contours is consistent with the X-ray
centroid thus confirming the symmetry of emission in the cluster. The
bright point-like source  seen in the contours in
Figure 7 at 1\farcm6 Southwest of the cluster center at
$\alpha = 04^{\rm h}43^{\rm m}05\fs8$ and
$\delta = $+$02\deg09\arcmin05\farcs5$ (J2000) is the QSO.

There are 768 $\pm$ 70 net counts in the cluster region within a circle
of 140\arcsec\ radius (583 h$_{50}^{-1}$ kpc at the distance of the
cluster) after subtraction of the emission due to the QSO. Assuming a
temperature of 5.5 keV, as determined from ASCA data, fractional
metallicity of 0.3 the solar value, and hydrogen column density along
the line of sight N$_{H}$=9.12$\times$10$^{20}$ cm$^{-2}$, we determine
a flux in the 0.1$-$2.4 keV energy band of (1.57$\pm$0.14) $\times$
10$^{-12}$ erg cm$^{-2}$ s$^{-1}$. The X-ray luminosity in the same
energy band is L$_{x}$=(2.85$\pm$0.25)$\times$10$^{44}$ h$_{50}^{-2}$
erg s$^{-1}$. The point-like source identified with the QSO has
136$\pm$16 net counts in a 28\arcsec\ radius circle. Assuming a power
law spectrum with an energy index $\alpha$=1.0 and the same N$_{H}$ as
before, the flux in the 0.1$-$2.4 keV band is (4.2$\pm$0.5) $\times$
10$^{-13}$ erg cm$^{-2}$ s$^{-1}$. The corresponding QSO X-ray
luminosity is
(4.8$\pm$0.6)$\times$10$^{44}$ h$_{50}^{-2}$ erg s$^{-1}$.

\subsection{\bf X-ray Profile}

A radial profile of the X-ray emission is obtained by summing the HRI
counts in concentric annuli of width 2\arcsec\ centered on the peak of
the X-ray emission and dividing by the area of the annuli.  This is
possible out to a radius of 140\arcsec\  (583 h$^{-1}_{50}$ kpc) at
which point the profile becomes indistinguishable, within the
noise, from the background level.  The profile is then fit with a
$\beta$ model (Cavaliere and Fusco-Femiano, 1976) described by
the standard form $S(r) = S_0(1+r^2/r_c^2)^{-3\beta+1/2}$, with
S$_{0}$~= central surface brightness, $r_{c}$ = core radius
and $\beta$~=~slope parameter. The values of the best fit  are
S$_{0}$~= 2.2 $\times 10^{-12}$ erg cm$^{-2}$ s$^{-1}$ arcmin$^{-2}$,
$r_c = 8\farcs0\pm1\farcs1$
and  $\beta = 0.46\pm0.03$. Figure 8 shows this fit to
be  well within the errors. Using the HRI point spread function
described in the Rosat User's Handbook, we deconvolved an
azimuthally symmetric image generated by revolving the fit.
A best fit of the $\beta$ model to the profile of the
deconvolved image has parameters $r_c = 6\farcs4\pm1\farcs1$
and $\beta = 0.45\pm0.03$. At the distance of the cluster this core
radius corresponds to 26.7 h$_{50}^{-1}$ kpc.
The size of the core radius could indicate the presence of a
cooling flow. However, there is no evidence for excess
emission in the very core of the cluster, thus any cooling flow
must be of marginal significance.
Small core radii (less than 40--50 kpc) and small $\beta$ values
($\sim$ 0.5), such as this cluster seems to
have, have been observed recently in other clusters; see for instance
MS1512$+$3647 ($r_c = 6\farcs9\pm1\farcs1$, $\beta = 0.524\pm0.031$;
Hamana \etal 1997) or our own unpublished HRI data for
MS2137-2353. The small size of the core radii and  $\beta$s for these X-ray
selected clusters is an issue that will be investigated in
future work.

\section{Results}

\subsection{Light and gas distribution}

The optical morphology of \ms04 is unusual. It is a poor cluster with a
luminous and compact core (R=14.8 within 24\arcsec) characterized by the
presence of several bright galaxies and numerous fainter ones, all
embedded in the low surface brightness halo. As suggested earlier
(Luppino \etal 1993), we may be seeing a cD galaxy in the act of
cannibalism. We have applied a gaussian filter with comparable $\sigma$
(2\arcsec\ in the HST image and 4\arcsec\ in the X-ray image) to
compare the light and gas distribution of the cluster core.
The X-ray centroid is within 1\farcs5 from the peak of the light
distribution: both distributions have almost circular symmetry with
similar ellipticity values ($\epsilon = 0.17$ in optical vs.
$\epsilon = 0.15$ in X-rays). We now proceed to determine the mass of
the cluster from each of three techniques, and compare one with the other.

\subsection{Mass estimate: Virial}

An estimate of the cluster mass is first attempted using the optical
data, via application of the virial theorem. From the virial theorem
equation $ M = 3 R_{v} \sigma_{los}^{2} /G $ and using the  measured
$\sigma_{los} = 872$ km s$^{-1}$, a mass of
4.8$^{+1.5}_{-0.94}\times10^{14}$ h$_{50}^{-1} M_{\sun}$ is obtained
for a value of the
three-dimensional virial radius R$_{v}$=0.91 h$_{50}^{-1}$ Mpc
(R$_{v}=(\pi/2)R_H$ where R$_H$ is the projected harmonic mean
radius). This mass is defined by the radial extent of the region
sampled; it is an underestimate of the total mass if the cluster extends
beyond the size of the observed field (in our case $\sim$ 1.5
h$_{50}^{-1}$ Mpc). We remind the reader that the accuracy of this
method depends on the assumptions that the velocity dispersion is
isotropic and that the cluster is no longer undergoing net expansion
or contraction.

\subsection{Mass estimate: X-ray}

Although there is no concern about isotropy of the X-ray emitting gas,
the mass determined using X-ray data does depend on assumptions
involving spherical symmetry and hydrostatic equilibrium. These
assumptions are found to be valid on the average
in N-body simulations studies of cluster formation by
Schindler \etal (1996) and Evrard \etal (1996). \ms04 does not have
any obvious substructure or strong shock wave that could affect the
mass determination. The cluster seems a relatively undisturbed cluster
(see Figure 7). The presence of small bumps may be
caused by some motion of collisionless matter and of intracluster gas.
Allen \etal (1996) caution  against the use of single-phase analyses
with ROSAT data, since the presence of a cooling flow with distributed
mass deposition implies that the central Intra Cluster Medium (ICM)
has a range of temperatures and densities at any particular radius:
i.e. the ICM is multi-phase. There is no evidence of any significant
cooling flow in \ms04, thus we feel that the assumptions adopted for
the deprojection outlined below are largely correct. In any case it is
not possible to resolve this cluster with  {\it ASCA}, or try to model
the temperature as a function of distance from the center of the
cluster since there are not enough photons for good limits. Thus in
the following we will assume that no temperature gradient is present.

 From the HRI surface brightness profile, assuming a constant
temperature, we have derived the three-dimensional density distribution
of the gas from the two-dimensional image by following the deprojection
technique of Arnaud (1988).  The data have been binned this time
in annuli of variable step (see plus signs in Figure 9 and Figure 10)
to have enough counts in each bin and thus reduce the statistical
uncertainty of the derived parameters. The integrated mass in
all forms can then be derived as a function of radius directly from
the equation of the hydrostatic equilibrium
\begin{equation}
M(r) = - {r  k T_g \over G \mu m_p} \left[ {d\ln \rho_g \over d\ln r}
+ {d\ln T_g \over d\ln r}\right],
\label{mr}
\end{equation}
\noindent where the symbols have
the standard notation. The radial dependence of the gas density
$\rho_g$ is given by the ROSAT HRI observations. $T_g$ is
the intra-cluster temperature, $\mu$ is the mean molecular weight of
the gas, and $m_p$ is the proton mass. The constant intra-cluster
gas temperature of  5.5 keV measured by ASCA is assumed.
Our estimates for the gas and gravitational mass within 140\arcsec\
($\sim$600 h$_{50}^{-1}$ kpc at the cluster redshift) are
(3.1$\pm0.25)\times10^{13}$ h$_{50}^{5/2} M_{\sun}$ and
(1.3$\pm0.2)\times10^{14}$ h$_{50}^{-1} M_{\sun}$.
The corresponding gas mass fraction,
M$_{gas}$/M$_{tot}$=(23$\pm$0.03)\% h$_{50}^{3/2}$, is typical
of the inner regions of rich clusters (David \etal 1995) and
consistent with that of low redshift clusters (White and Fabian 1995).

We have also deprojected the finely binned X-ray profile
of Figure 8 assuming the standard $\beta$ model.
Our version of this analysis follows. The volume emissivity,
$\epsilon_{v}$, is found from a deprojection of the profile.  For
the standard functional form, the deprojection yields $\epsilon_v \propto (1 +
r^2/r_c^2)^{-3\beta}$.  The emissivity is proportional to the density
squared times the cooling function.  If we assume the gas is nearly isothermal,
the density is $\rho_g = \rho_g(0)(1 + r^2/r_c^2)^{-3\beta/2}$.
The gas density distribution is presented, in Figure 9, in units of
protons cm$^{-3}$. However, the normalization is irrelevant to
all that follows. For the purpose of assessing various sources of
error, we continue the analysis of the fit to the raw data as well as
the fit to the deconvolved data, but, in the end, quote results from
the deconvolved data.

If one substitutes the $\beta$ model law into Equation~1,
and assumes constant temperature gas, or at least that the density
falls off much more rapidly than the temperature, then one finds the
following expression for the underlying mass as a function of radius:
\begin{equation}
M(r) = {3 \beta r_c \sigma_g^2 \over G} {s^3 \over 1 + s^2},
\end{equation}
where $\sigma_g = \sqrt{kT_g/\mu m_{\rm p}}$ is the 1-d velocity dispersion
of the gas, and $s \equiv r/r_c$. The $\beta$ model mass distribution for
\ms04 is shown as the dashed line (fit to deconvolved data) and dotted
line (fit to raw data) in Figure 10.

The density distribution  of the $\beta$ model is easily evaluated to be
of the following form:
\begin{equation}
\rho^T = {1 \over 4 \pi r^2} {d M(r) \over d r} = {\rho^T(0) \over 3}
	{3 + s^2 \over (1 + s^2)^2},
\end{equation}
where $\rho^T(0) = 9 \beta \sigma_g^2 / 4 \pi G r_c^2$.  This profile
resembles an isothermal mass distribution at small radii, deviates by a
maximum of 1.81 at 11$r_c$ and asymptotically approaches 1.5 times the
isothermal distribution at large radii.

To compare to the gravitational lens results, one needs the projected
cumulative mass profile.  The surface mass in the $\beta$ model
is given by
\begin{equation}
\Sigma(R) = 2\int_0^\infty \rho^T dz = {\pi \over 3} \rho^T(0) r_c
	{2 + S^2  \over (1+S^2)^{3/2}},
\end{equation}
where $z$ is the distance along the line-of-sight, $S = R/r_c$, and R refers
to the radius in the plane of the sky.  This, as it turns out, is
exactly the same form as the surface density derived from the
``isothermal" case ($\alpha = 1/2$) of the Blandford-Kochanek
formula of the lensing potential (Blandford and Kochanek, 1987).
The normalizations are also equal by setting
$\sigma_{BK} = \sqrt{3\beta/2}~~\sigma_g$.

The projected cumulative mass profile is given by
\begin{equation}
	M(R) = 2 \pi \int \Sigma(R) R dR =
	{2 \pi^2 \over 3} \rho^T(0)r_c^3 \sqrt{1+S^2}\left[1 - {1 \over
	1+S^2}\right],
\end{equation}
\noindent and is shown as a dash-dot-dot-dot line  in Figure 10.
In the region of the observed multiple arc systems the projected mass
is a few times the unprojected mass.  At radii $>> r_c$,
the projected mass eventually drops to $\pi/2$ times the
unprojected mass at a similar radius.

We also examine an isothermal
distribution for the underlying mass.   A numerical solution to the
differential equation of hydrostatic equilibrium is found for the case
of constant temperature.   A finely spaced solution of the density,
$\rho_{iso}(s)$,  is integrated to give $M(s)$ and $\int M(s)/s^2 ds$.
The gas density distribution can be solved for in
Equation~1:

\begin{equation}
	\rho_g(r) = \rho_g(0) {\rm exp}\left[-{G \over \sigma_g^2 r_c}
\int_0^s {M(s') \over s'^2}ds'\right].
\label{Eqintm}
\end{equation}

We then square the gas density distribution, using the same core radius
as in  the $\beta$ model,  and project it onto the plane of the sky to
compare with the observed deconvolved surface brightness.

Using the asymptotic solution at large radius for the isothermal sphere,
an approximate value for the central density can be derived that matches
the $\beta$ model fit to the surface profile,
\begin{equation}
	\rho_{iso}(0) = {27 \beta  \sigma_g^2 \over 8 \pi G r_c^2}
\end{equation}

However, a better overall fit is found with $\rho_{iso}(0)$ at 0.77
times this value, as is shown in Figure 11.  Even with some freedom
in choosing $r_c$ and $\rho_0$, the isothermal sphere model does not
adequately fit at all radii.  It appears that the simpler $\beta$ model
is, in fact, a better representation of the underlying mass distribution
than an isothermal sphere.

\subsection{Mass limits: Lensing}

A reconstruction of the arc images to background source galaxies was
attempted using the morphological information in the HST images.
As with other previously known lensing clusters (Kneib \etal
1996; Smail \etal 1997 among others) the high resolution of WFPC2 reveals
significantly more lensed features than by using ground-based
telescopes. Several arcs in \ms04 are resolved by the HST
into bright knots. Others like arc A1 which appeared as the most
prominent arc from the ground data, is actually a distorted image of a
galaxy but not necessarily a strong lensed object.
We will follow and extend the nomenclature in Luppino \etal (1993) for
both the galaxy names and the arcs. However, with the HST
resolution available,  we now think that objects labelled as
A10, A14 and A15 are not strong arcs and they are not labelled in
Figure 2. Arc A11, which may be an arc, is out of the
field of view of this image.
A solution is obtained assuming a flattened potential
(Blandford and Kochanek, 1987),

\begin{equation}
	\phi = {\pi \sigma_{BK}^2 r_c \over \alpha G}
	(1 + (1-\epsilon){\rm x^2} + ( 1 + \epsilon){\rm y^2})^{-\alpha}
 \end{equation}
where $r_c$ is the angular extent of the core radius, x is the run along
the major axis in units of $r_c$, and y is the run along the minor axis.
A position angle to align coordinates with the major axis must also be
specified.  The parameters $\sigma_{BK}$, $r_c$, and $\epsilon$ are simply
fitting parameters and are  not easily converted to any real physical
quantities except in the  $\alpha=0.5$ case.

The observed position in the plane of the sky
$X_{\alpha,\delta}^o$ is
related to the source position $X_{\alpha,\delta}^s$ by,
\begin{equation}
{\bf X}_{\alpha,\delta}^o  =
	{\bf X}_{\alpha,\delta}^s - \nabla_{x,y} \phi d_{ls}/d_s
\end{equation}
where $d_s$ is the comoving angular diameter distance of the source and
$d_{ls}$ is the comoving angular diameter distance of the source as seen
by the lensing cluster (Hammer and Nottale, 1986; Peebles, 1993).

The full set of free parameters is: (x,y) position of center, position
angle on the sky, $r_c$, $\alpha$, $\epsilon$, $\sigma_{BK}$, and
the redshifts of all source objects.  We do not restrict the model
by the observed light distribution. In the end, the light and matter
can be compared and information can be gleaned as to the
degree to which the potential follows the light distribution.

The main difficulty stems from properly identifying which sets of arcs
are to be associated as being counter images of the same source. The
details of the model will be presented in Shaya  \etal (1998).  Here we
present the results obtained for the mass determination. Since we do not
have a redshift for any multi-arc system, we can obtain only upper limits
to the mass and lower limits to the redshift.  An upper limit to the
mass derives from the fact that no credible counter image is found for
Arc 1, the arc with known redshift.
A lower limit ultimately derives from the
fact that the term  $d_{ls}/d_s$, in standard cosmologies with
$\Omega$ = 1.0 (0.1), approaches the finite value of 0.91 (0.84)
as the redshift of the source goes to infinity.

Crude estimates for the values of the model parameters were first
established by solving for the case that the pattern of 4 knots in A5
corresponds to the  4 knots in A6.  With this model in hand one could
explore different source plane distances to see what other arcs are
counter images of each other.  When another was found, a $\chi^2$
minimization program to bring the knots of both systems together in the
source plane provided more highly constrained values on the model
parameters.  This procedure was repeated, with more terms in the
$\chi^2$ each round, until no new set of counter images could be seen.
If an incorrect association is made between arcs, then poor values for the
parameters are assumed and this will lead to a dead end, in the sense
that no new systems will be found.  One must then back up and try a
different pairing that will lead to further progress.

The following sets of arcs were found to be associated with their own
source object with the range in redshifts set by the two limiting
models: Arcs 5 and 6 (0.60 $<$ z $<$ 1.6), Arcs 8, 9, 12 and 24 (0.53
$<$ z $<$ 1.1),  radial Arc 17 and Arc 18 (0.59 $<$ z $<$1.5), Arc 7 and
radial Arc 16 (0.59 $<$ z $<$ 1.5), and Arcs 2, 3, 20 and the faint, extreme
Southern extension of A9 (0.75 $<$ z $<$ $\infty$).
A best solution that solves for
these arcs simultaneously requires values for $\epsilon =
0.074\pm0.005$, and for $r_c = 1\farcs55\pm0.01$
corresponding to 6.4 h$_{50}^{-1}$ kpc, a position angle of
$74\fdg2\pm1\fdg4$, and  $\alpha =0.760\pm0.007$. However, these are
uncorrelated errors.  For correlated errors in which all of the other
parameters are permitted to change as each parameter is tested, the
errors in position angle and  ellipticity are  $2\fdg4$ and 0.03,
respectively.  The core radius could be varied from 0 to 3\farcs35
(within 1 $\sigma$ in $\chi^2$).  The value for $\alpha$ could be varied
between 0.74 and 0.97 with little penalty in $\chi^2$ except at the
edges of this range.

The ellipticity in the lensing model reflects the ellipticity of the
potential.  It is expected to be about one third of the ellipticity of
the underlying mass distribution (Mellier \etal 1993).  Here, we find
a ratio between the light distribution and the potential ellipticities
of 2:1.  It appears, therefore, that the value of the ellipticity of the
light distribution is intermediate between that of the potential and the mass.

Arcs 5 and 6 identified in Figure 2 provide an important
constraint on the mass geometry because a single source, with consistent
complex structure, explains both. We failed to associate Arc 1, the only
arc with known redshift at z$=$0.5317, with any counter image.  If this
is, indeed, because there is only one image, it provides an upper limit
to the mass.  For the maximal mass model, Arc 1 is allowed to be just
beginning to form a counter image.  At this point, the source image of
Arc 1 is becoming alarmingly distended in a direction pointing toward
the center.  Thus a potential greater than the maximal mass model
is unlikely, based solely on the form of the source
image of Arc 1.

\section{Discussion}

The projected mass as a function of radius for X-ray data and lens
modelling  is presented in Figure 12. Out to the radius of 40
kpc, the X-ray mass model has just barely enough mass to be consistent
with the lens mass.  By 100 kpc, the outermost radius with
strong lensing, the X-ray mass model appears to fall about a factor
of 1.5 to 2 too low.
The errors in the X-ray determined mass, are $\sim 24\%$ (90\%
confidence) after folding in errors in ASCA determined temperature with
uncertainty in individual bin counts.  Therefore, perhaps our model is
too simple. The model that we describe below is presented as a
possible  way to reconcile the X-ray and lensing mass. Just adding a
little more complexity to the model, and one that is well motivated by
the complex structures observed on large scales, the  X-ray and lens
mass determinations can be reconciled. There may be other similarly
complex models that resolve the mass discrepancy. Here we prove only the
existence of a solution, not uniqueness.

It needs to be noted that the path followed by light from background
sources is affected by the mass of the entire column through the
line of sight.  It may be that the projection of just the cluster
does not fully represent the total mass in the column.
As a next simplest model, we look for a mass
distribution made up of two isothermal spheres centered on the cluster
which distributes the hot gas in such a way so that its surface
brightness remains consistent with the observed X-ray brightness
profile, but has a projected mass consistent with the gravitational lens
results.  It is true that we are extrapolating beyond the last point at
which the gas could have been heated sufficiently to radiate X-rays.
We are trying to reconcile here the disparity between X-ray determined
mass and gravitational lens mass plus we are trying to understand the
steep slope of the gravitational lens mass.  Both problems
are best dealt with by the proposed theory that clusters are embedded in
superclusters. It is exactly the lack of information in the X-ray gas
at large radii which we are exploiting to find a single model that fits
both the X-ray and the gravitational lens.
A solution is found with one isothermal sphere of
core radius unchanged from the previous analysis but with a second
sphere with core radius between 30 -- 50 times larger.  For definiteness,
we show a solution with 40 times larger core radius, $r_c = 1.06$ Mpc.
The second component sphere has a central density 150 times less than
the first component.  The velocity dispersion of the second component is
then $40/\sqrt{150} = 3.26$ times that of the first component, which put
it at nearly  $\sigma_v = 2000$  km s$^{-1}$. It is unclear, however, whether
this component should be attributed to a second, warmer dark matter
particle,  or (more likely) to late fall into the cluster.
Figure 13  compares the new density distribution (dash-dots)
with the two from the previous sections, the $\beta$ model (dotted) and
the single isothermal sphere (dashed).    The mass enclosed at each
radius is presented in Figure 14.  Although we continue the
distribution out to 10 Mpc where it reaches a total mass of $2\times
10^{16} M_{\sun}$, the distribution could start falling off at a few Mpc
with little effect in what follows.  The hump in the density and mass
distributions beginning at a few 100 kpc could simply be a
representation of the supercluster within which the cluster resides.  In
fact the total mass within 10 Mpc is quite reasonable for a major
supercluster such as the Coma Supercluster or the Great Attractor.

We again use Equation~6 to calculate the expected X-ray
profile and to set the normalization of $\rho^T$.  This time the fit
(Figure 15) fits well over most of the range but, admittedly,
is a little low at 300 kpc, but it is nonetheless acceptable.  The
signal-to-noise is low at these radii and many of the  bins give only
upper limits.  A slight error in background subtraction may contribute
to the small discrepancies.
Finally, the dash-dot line in Figure 12 shows the projection of
the 2-isothermal spheres model and it appears to just fit the minimum
mass model both in terms of slope and amplitude.  The explanation then
for the discrepancy between the X-ray determined mass and the
gravitational lens mass might simply be the fact that the X-ray mass is not
sensitive to the larger scale structure within which the cluster is
embedded. We have not yet explored how sensitive the results are to
coalignment of these two potentials.  In future work, we will
examine how stringent is the requirement for the cluster to be at the
center of the larger scale structure.

Other alternatives for the cause of discrepancy include the
possibility of temperature variations in the X-ray gas. We will have
to wait for better X-ray spectroscopy with {\it AXAF} since {\it ASCA}
cannot resolve this cluster.  Bartelmann and Steinmetz (1996) suggested
that the presence of substructure and line-of-sight alignments of
material towards the cluster core may contribute to the discrepancy
observed, since they will increase the probability of detecting
gravitational arcs in the clusters and thus enhance the lensing masses,
without significantly affecting the X-ray data.
The mass model prediction presented here will be tested by the weak lensing
modelling which is underway by members of our team.  The details may disagree
because the supercluster is not expected to be spherically symmetric.
The strong lens model examined in this paper is sensitive to the distribution
along the line of sight to the center while the weak lensing will examine
the distribution at large distances in the other two dimensions.

\section{Conclusions}

The observational data from a multiwavelength study of the cluster \ms04
have been presented, together with the analysis of the mass
distribution as obtained by several techniques.
For the HST/WFPC2 image, we focused on modelling of the gravitational
lensed arcs distributed in the inner $24\arcsec$ radius.
Ground based telescopes were used to obtain the
velocity dispersion of the galaxies in the cluster and determine the
virial mass.  We used {\it ASCA} data to derive a temperature for the X-ray
gas.  A {\it ROSAT/HRI} image was used to derive an emission profile which is
analyzed, assuming that the hot gas is in hydrostatic equilibrium and in a
spherical potential, to derive the form of the cluster potential.

 From possible multiple images formed by gravitational lensing of 5
background sources, we have derived limits to the mass distribution
in the range 50 -- 100 h$_{50}^{-1}$ kpc in \ms04.  For the central
24\arcsec\ (100 h$_{50}^{-1}$ kpc) region encircled by the arcs, the
possible range in projected mass is
$6.6 - 9.5\times10^{13}$ h$_{50}^{-1} M_{\sun}$.
The mass profile appears to grow with radius considerably more rapidly
than an isothermal model or a $\beta$ model.  We have also used X-ray
data to obtain a mass distribution from the inner few kpc out to
nearly 600 h$_{50}^{-1}$ kpc.  There is no evidence for excess
emission in the very core of the cluster which could indicate
the existence of a cooling flow, thus justifying the assumption of
constant temperature for the deprojection technique. The mass
distribution obtained is well fit by a $\beta$ model, described here,
but not well fit by a single isothermal distribution.  The X-ray
derived projected mass profile is below the lensing mass profile.  At
50 h$_{50}^{-1}$ kpc, it is 20\% below, which is just within the
errors, but by 100 h$_{50}^{-1}$ kpc it is a factor of 2 below.
However, more general models fit the data better and allow a wider
range of masses that are consistent with the lensing data.

The virial mass derived from the galaxy velocity dispersion,
$4.8^{+1.5}_{-0.94} \times 10^{14}$ h$_{50}^{-1} M_{\sun}$  is
intermediate between the extrapolations to  900 h$_{50}^{-1}$ kpc
of the other two  profiles. As in other cases reported in the literature
we find discrepancy between the X-ray and lensing estimates.

We tentatively explore the possibility of reconciling these mass
estimates with a mass profile that increases more rapidly than the
X-ray $\beta$ model at large radii.  The model explored is one with
two isothermal spheres; one has a core radius of 26.7 h$_{50}^{-1}$
kpc and  the other core radius is 1 h$_{50}^{-1}$ Mpc.  The
central densities for the two components have a ratio of 150:1.  With
these parameters, a fit to the X-ray profile is reasonable and the
projected mass profile is consistent with the minimum mass model for the
gravitationally lensed arcs. The total mass out to 10 Mpc required by
this model is about $2 \times 10^{16}$ h$_{50}^{-1} M_{\sun}$, which
could indicate the existence of a supercluster  of galaxies with a mass
comparable to the Coma Supercluster or the Great Attractor.
Other alternatives for the discrepancy of the mass estimates
include: a temperature gradient in the X-ray  gas that may conspire
against the models used; or the lensing mass could be higher because
of line of sight projection effects. The asymmetric velocity
distribution of the galaxies and its low end extension could be an
indication that the mass of the cluster is not spherically distributed.

Similarly to other investigators we have found a discrepancy
between the X-ray and lensing mass determination (see among others
Miralda-Escud\'e and  Babul, 1995; Kneib \etal 1995 or the
exhaustive list of references in Wu and Fang, 1996, 1997).
Differently from other lensing clusters with X-ray data, \ms04 does not
give evidence of any ongoing merger which could severely disturb the
intracluster gas (i.e. like A2218, Kneib \etal 1995; A370,
Mellier \etal 1994) and thus explain the discrepancy. The simple
model presented here could be tested by: 1) a much deeper X-ray map
showing more details of the behavior of the surface brightness
profile (it may provide evidence of a second, large scale component)
but a much larger field of view than that given by HRI would be necessary;
2) detection of a temperature gradient, possible with higher
resolution instruments as the ones which will be flown on AXAF;
3) weak lensing analysis to greater radii using the large
CCD mosaic camera with 8100$\times$8100 pixels (Metzger, Luppino and
Miyazaki, 1995) which will obtain wide field images of clusters at
relatively low redshift such as the one presented here.

\acknowledgments

It is a pleasure to thank A. Wolter, R. Della Ceca and E. Radice for
useful discussions and for comments on the X-ray data analysis.
B. Tully and P. Henry did a careful reading of the manuscript. An
anonymous referee provided many useful comments that helped us
clarify several areas of this paper. This work has received partial
financial support from NASA-STScI grant GO-5402.01-93A, NASA grants
NAG5-2594, NAG5-2914, NSF AST95-00515 and CNR ASI grants ASI94-RS-10
and ARS-96-13.

\clearpage
\setcounter{figure}{0}
\begin{figure}
\epsscale{.98}
\figcaption{The full WFPC2 image of \ms04 ($z=0.1965$) in the
F702W filter, sum of all exposures and cosmic ray cleaned.
Orientation is explained in Figure~2.
}
\label{image1}
\end{figure}

\clearpage
\begin{figure}
\epsscale{.9}
\figcaption{WFC3 image, with local median average subtracted, of the
central $51\farcs2\times51\farcs2$ (512 by 512 pixels) of \ms04 with
names given for each arc. The plus sign marks the center of mass.
The cross sign indicates the center of this isophote fitting.
The figure has to be rotated by 152.46
clockwise to have North up and East to the left.}
\label{image2}
\end{figure}

\clearpage
\begin{figure}
\epsscale{.8}
\plotone{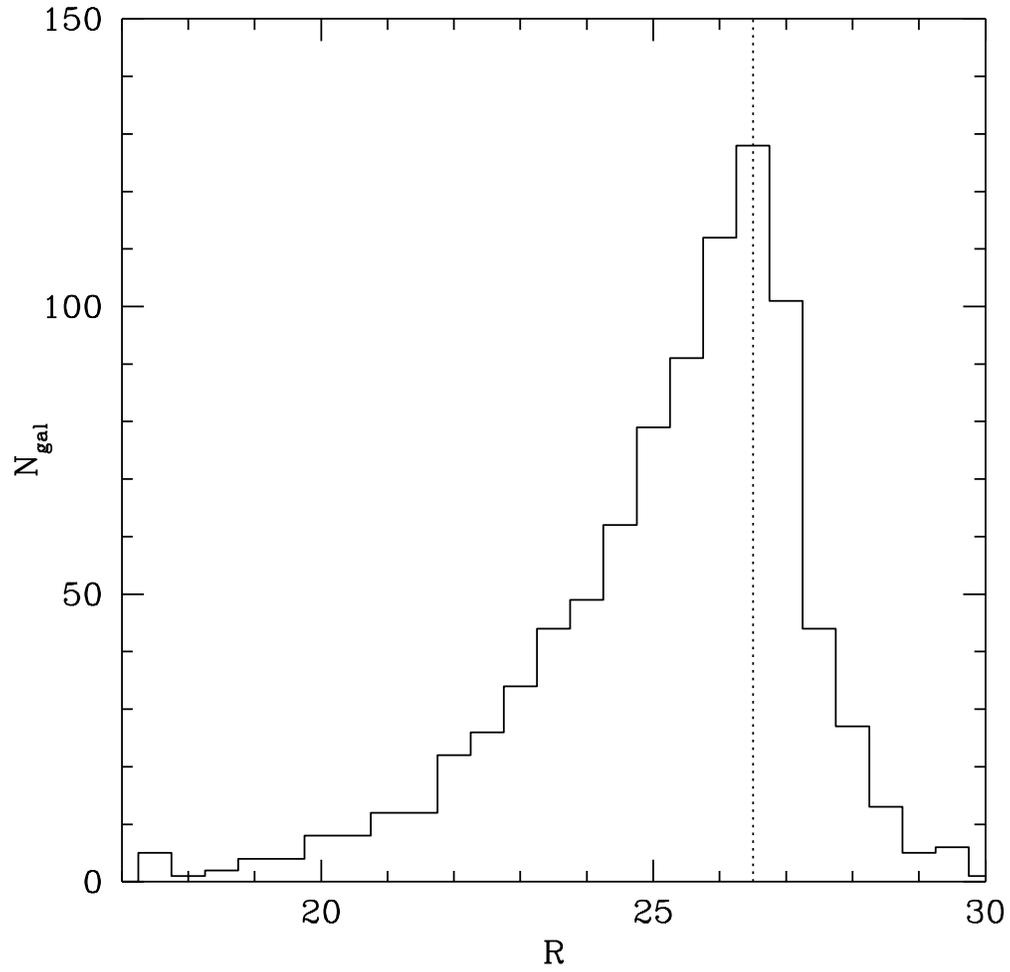}
\figcaption{The R band magnitude number counts. The decline at $R\geq26$
indicates that the counts are incomplete at fainter magnitudes.
}
\label{rcounts}
\end{figure}

\clearpage
\begin{figure}
\epsscale{.80}
\figcaption{B$+$R CCD image of \ms04 (adapted from Gioia and Luppino,
1994) showing objects with spectroscopic data. The field of
view is $3'.5\times 3'.5$, corresponding to 0.87 Mpc
$\times$ 0.87 Mpc at the redshift of the cluster (North is up and East
to the left).
}
\label{galaxies}
\end{figure}

\clearpage
\begin{figure}
\epsscale{.9}
\plotone{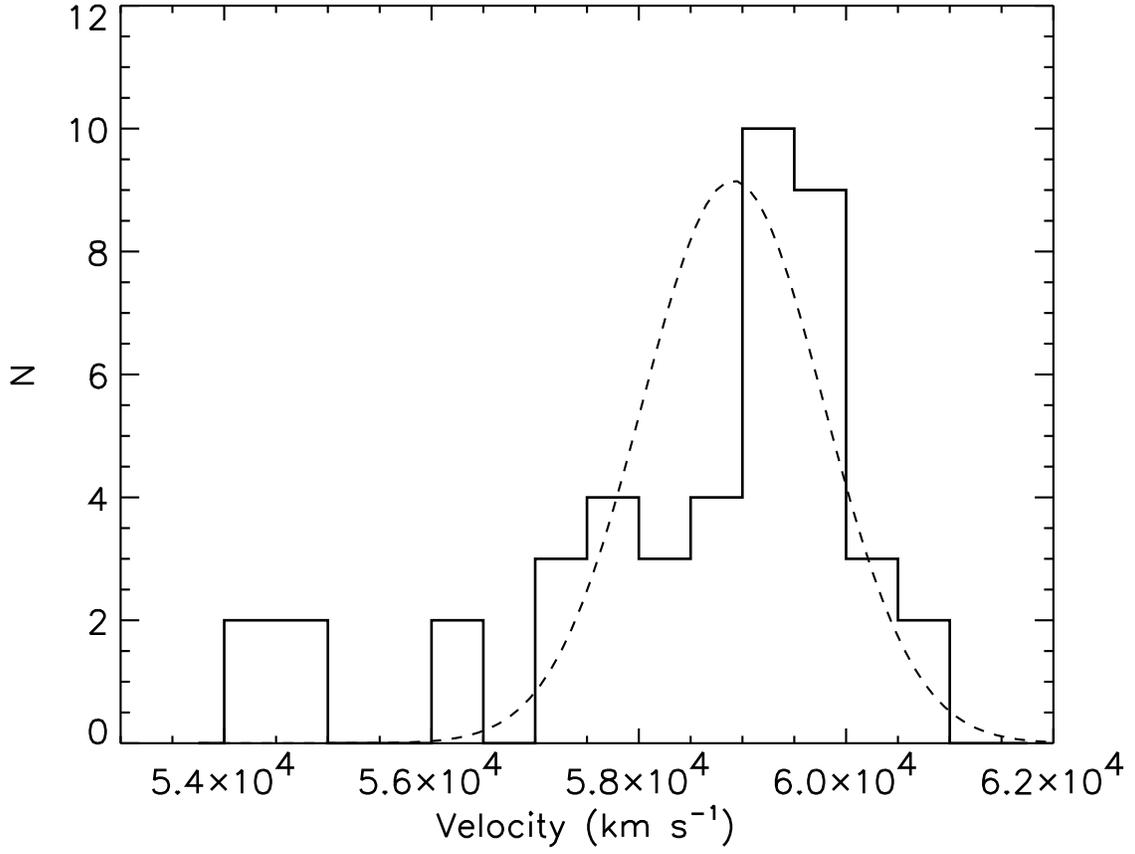}
\figcaption{Distribution of velocities for the entire sample of 44 galaxies
showing a low-velocity tail at 55,000 km s$^{-1}$.  The curve shows a
gaussian distribution with width given by the derived $\sigma_v$ and
normalized for the 40 galaxies considered cluster members.
}
\label{vhistogram}
\end{figure}

\begin{figure}
\epsscale{.95}
\plotone{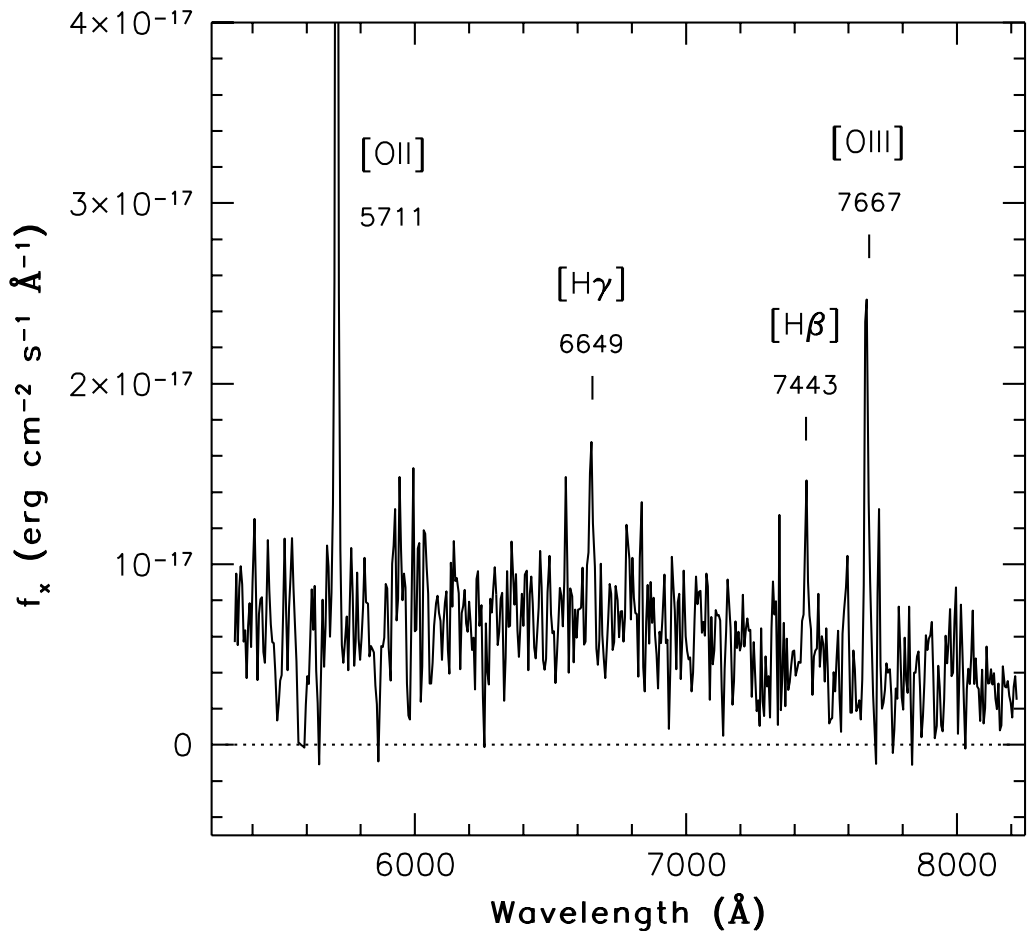}
\figcaption{CFHT optical spectrum of the distorted arc-like galaxy A1 to the
East of the cluster core. From O[II], H$\gamma$, O[III] and H$\beta$
emission lines a redshift of 0.5317 is obtained.
}
\label{A1spec}
\end{figure}

\clearpage
\begin{figure}
\epsscale{.9}
\figcaption{The HRI iso-intensity contour map overlaid on the optical CCD
frame of Figure~6. The X-ray image is smoothed with a
Gaussian with $\sigma=4\arcsec$. Isophotal contours are 0.044,
0.062, 0.086, 0.115, 0.156, 0.231, and 0.306 net counts arcsec$^{-2}$,
with the lowest contour being 40\% above the background of
0.10957$\times10^{-5}$ counts arcsec$^{-1}$. The point-like source
1\farcm6 to the Southwest is the QSO.
}
\label{hricontour}
\end{figure}

\clearpage
\begin{figure}
\epsscale{.9}
\plotone{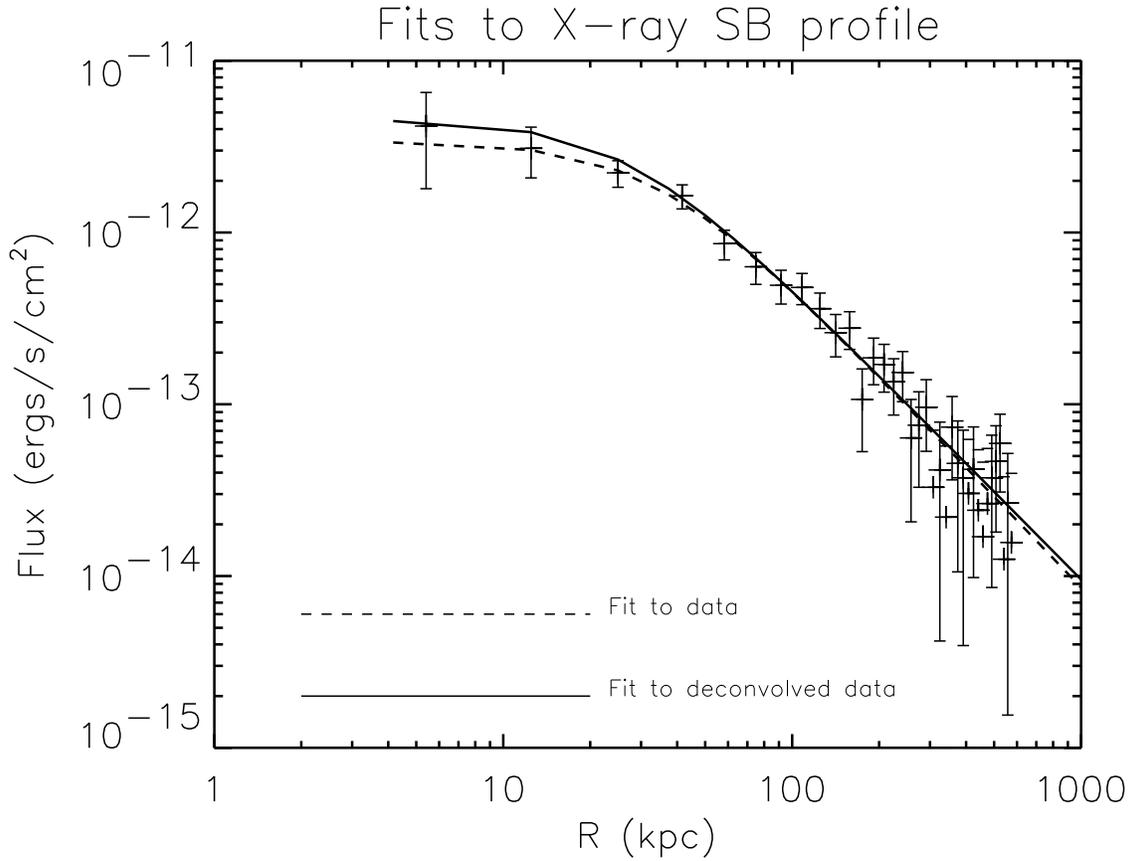}
\figcaption{The HRI brightness profile azimuthally averaged is presented
as plus signs with error bars.  The dashed line is the best fit of form
$S(r) = S_0(1+r^2/r_c^2)^{-3\beta+1/2}$.  The solid line is the best fit
of the same form to the deconvolution of the raw fit with the HRI point
spread function.  See text for values of fit parameters.
}
\label{xsb}
\end{figure}

\begin{figure}
\epsscale{.9}
\plotone{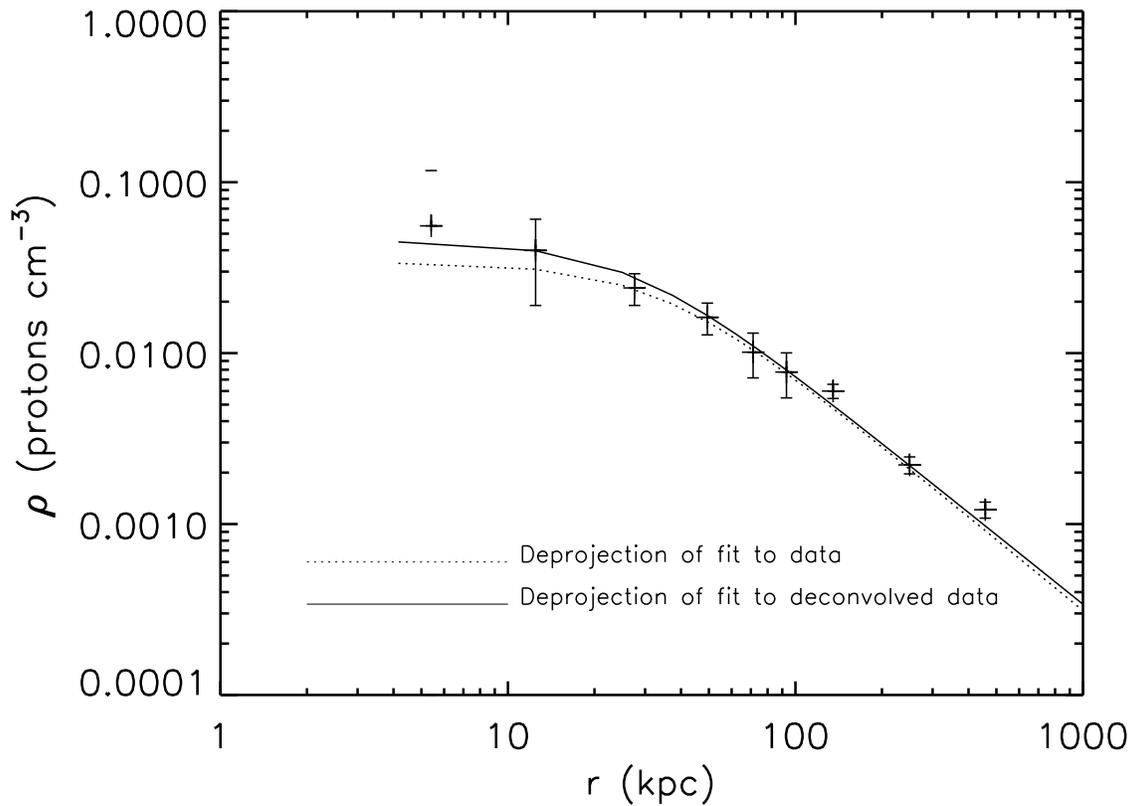}
\figcaption{Gas density profile derived from HRI brightness profile.   Plus
signs are derived from coarsely binned data. See text for details.
Dotted line is $\beta$
model fit, $\rho_g = \rho_g(0)(1 + r^2/r_c^2)^{-3\beta/2}$ to the raw
data.  Solid line is the same but for the deconvolved data.
}
\label{xrho}
\end{figure}

\begin{figure}
\epsscale{.85}
\plotone{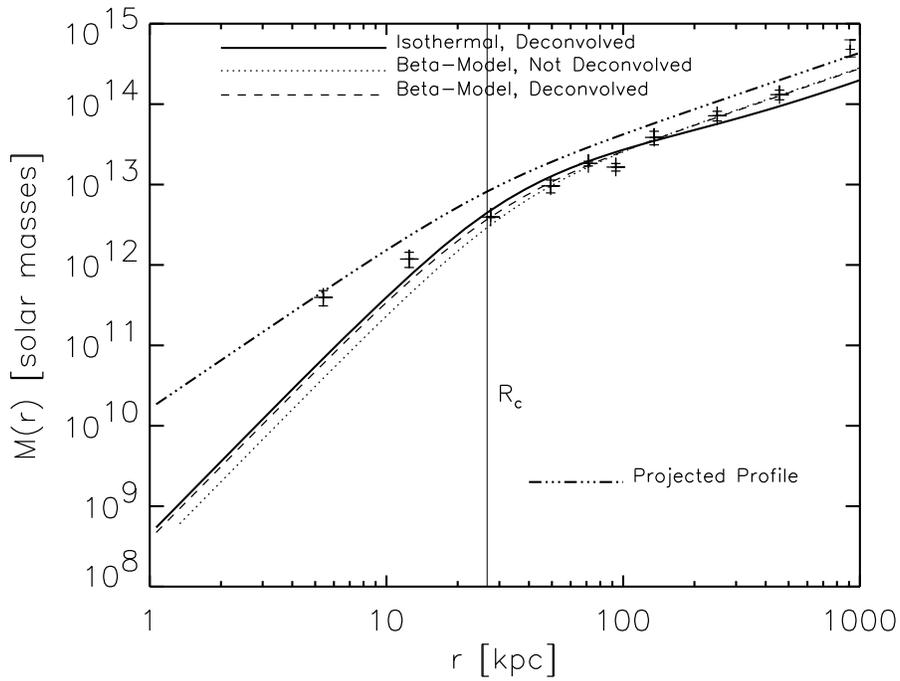}
\figcaption{Cumulative mass with radius as derived from HRI brightness
profile.  Plus signs with error bars are found by coarsely binning the
data and solving for M(r) independently for each bin. The dotted line is
a $\beta$ model fit to raw data. The dashed line is a $\beta$ model fit to
deconvolved data.  The solid line is from a model in which the total
mass is distributed as an isothermal sphere model. The thin lined
error bar at 900 h$_{50}^{-1}$ kpc shows the virial determination of
the mass. The dash-dot-dot-dot line is the projection of the $\beta$
model for comparison with gravitational lens mass determination.  A
vertical line is placed at the core radius of the fit.
}
\label{xmass}
\end{figure}

\begin{figure}
\epsscale{.9}
\plotone{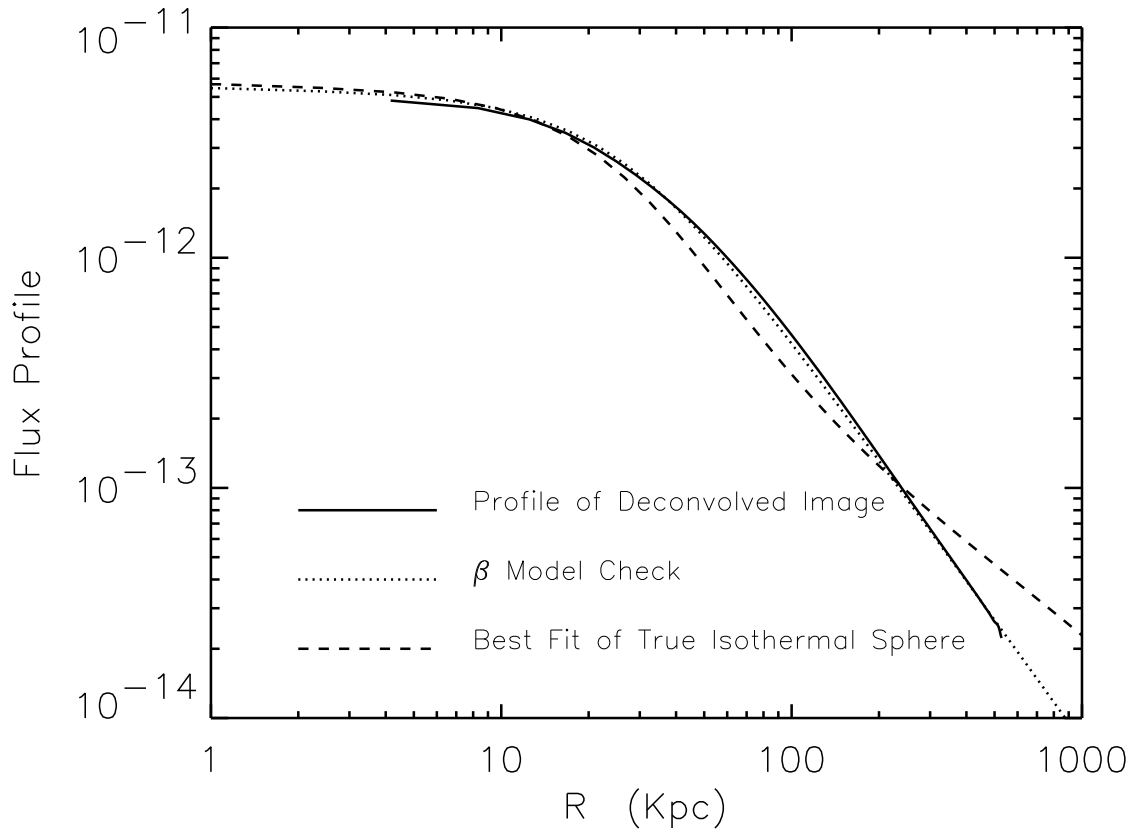}
\figcaption{Calculations of the brightness profile given the total mass
density distributions. See text for details. For the $\beta$ model
(dotted line), this amounts to reversing the procedure of deriving
M(r) and is thus just a check of the total procedure.  For the
model with the total mass following an isothermal distribution (dashed
line), the best normalization is determined by changing $\rho^T$ and
$r_c$ iteratively until a good fit to the X-ray brightness profile is
found. The $\beta$ model is a significantly better fit.
}
\label{iso}
\end{figure}

\begin{figure}
\epsscale{.85}
\plotone{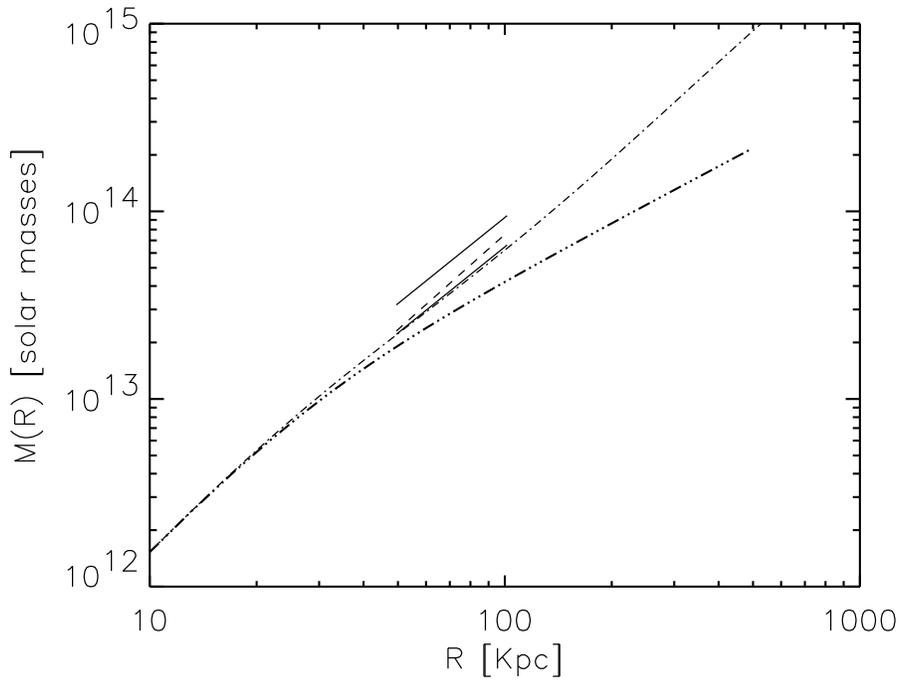}
\figcaption{Projected mass profile from lensing and X-ray data. The
dash-dot-dot-dot line is the $\beta$ model deprojection of the X-ray mass
profile.  From the gravitational lensing,  the two solid lines represent
the projected minimum and maximum mass models with $\alpha=0.76$.  The
minimum acceptable value for $\alpha$ is 0.74.  The dashed curve is
from the model with the maximum acceptable value for  $\alpha=0.97$.
The dash-dotted line is the projected mass of 2 isothermal spheres having
$r_c=26.7$ h$_{50}^{-1}$ kpc and 1 h$_{50}^{-1}$ Mpc and with
$\rho^T$ ratio of 1:150. }
\label{masses}
\end{figure}

\begin{figure}
\epsscale{.9}
\plotone{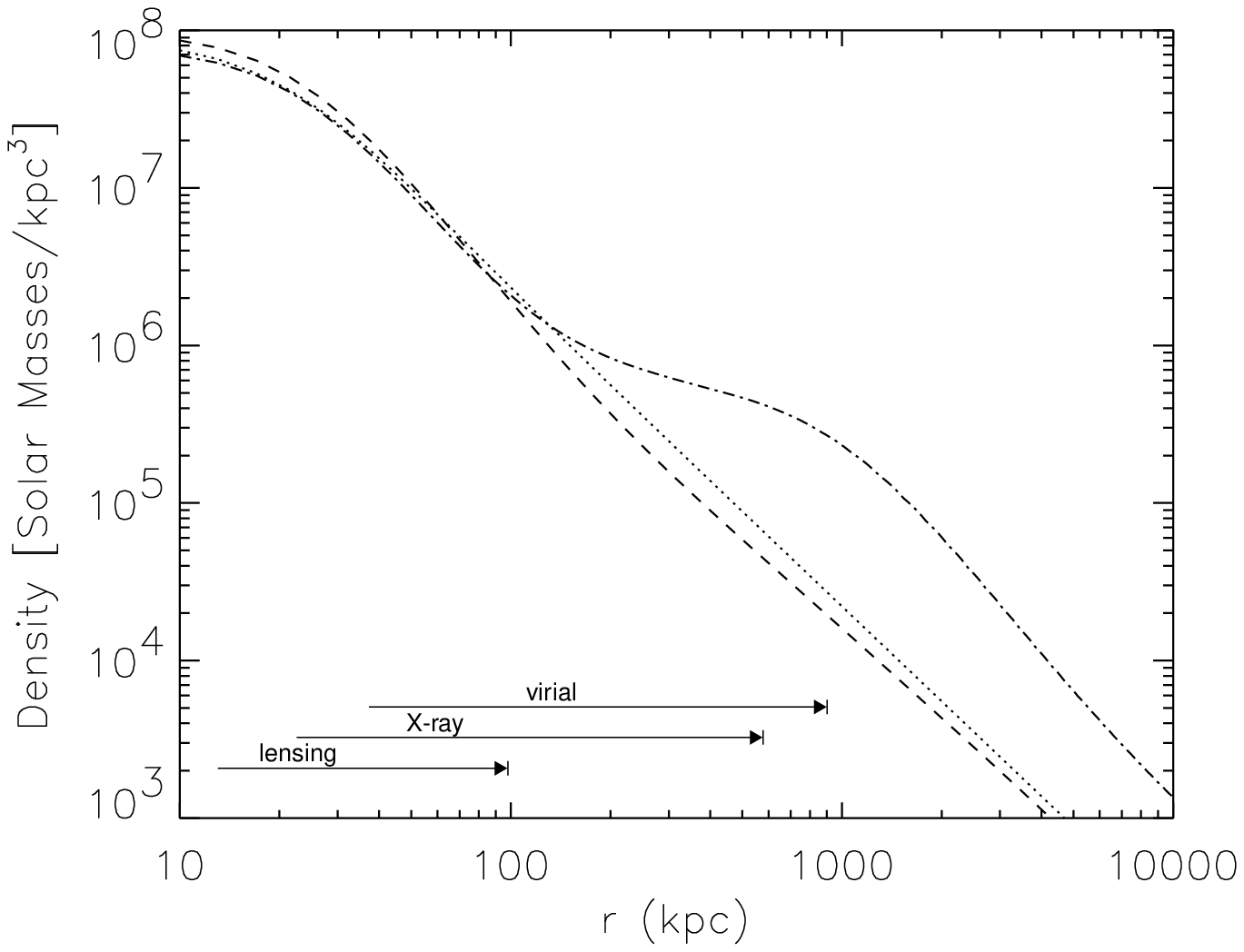}
\figcaption{Density Distributions.  2 concentric isothermal mass spheres
with $r_c=26.7$ h$_{50}^{-1}$ kpc and 1 h$_{50}^{-1}$ Mpc, central
density ratios 150:1 (dash-dotted line).  Also shown are the single
isothermal mass (dashed line) and the single $\beta$ model (dotted line)
used in the previous figures. The three thin solid lines parallel to the
x-axis indicate the ranges in radius constrained by the different kinds
of data.
}
\label{2isosrho}
\end{figure}

\begin{figure}
\epsscale{.9}
\plotone{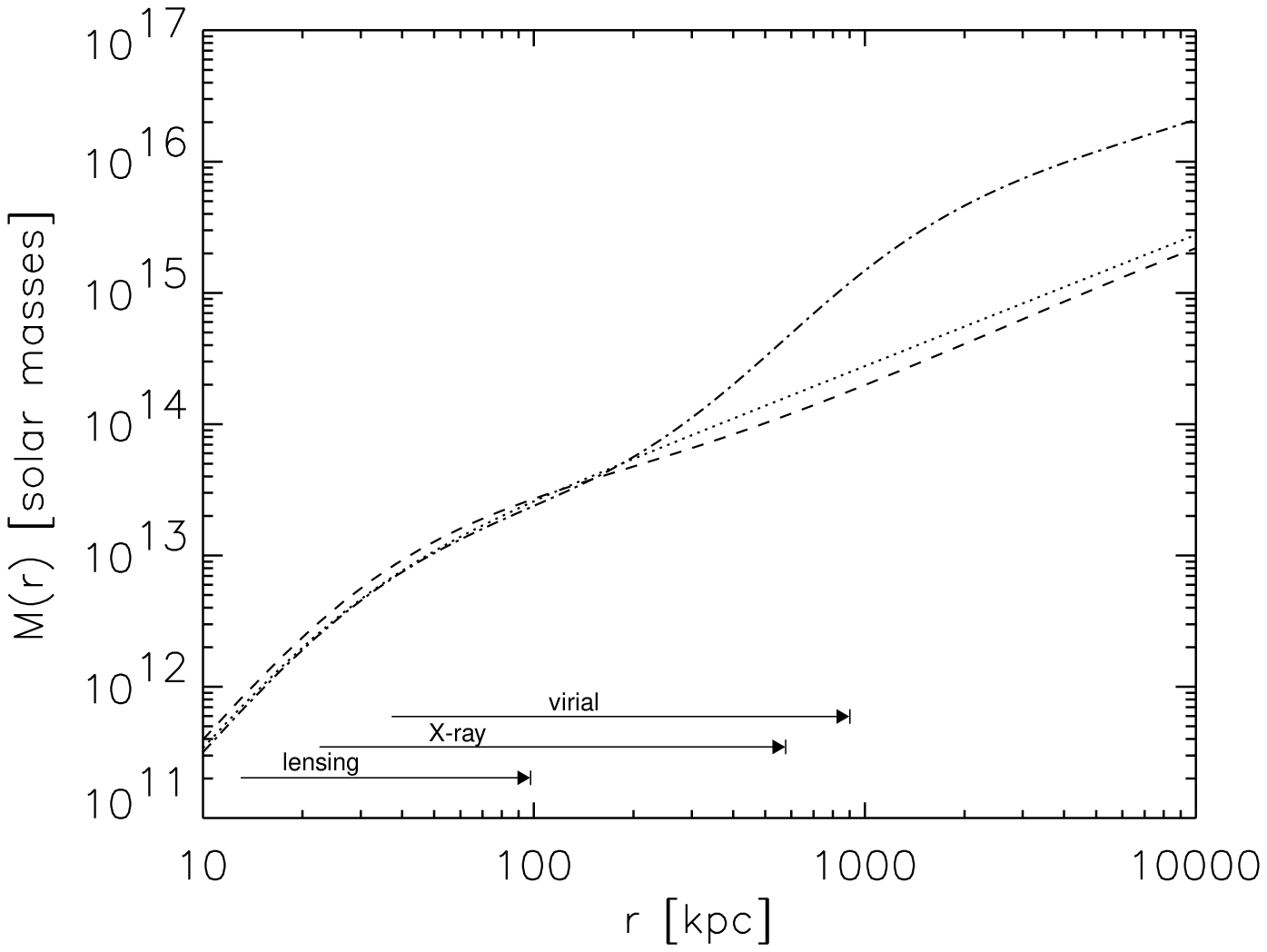}
\figcaption{Mass Distributions.  2 concentric isothermal mass spheres
with $r_c=26.7$ h$_{50}^{-1}$ kpc and 1 h$_{50}^{-1}$ Mpc, central
density  ratios 150:1 (dash-dotted line). Also shown are the single
isothermal mass (dashed line) and the single $\beta$ model (dotted line)
used in the previous figures. The three thin solid lines parallel to the
x-axis indicate the ranges in radius constrained by the different kinds
of data.
}
\label{2isosmass}
\end{figure}

\begin{figure}
\epsscale{.9}
\plotone{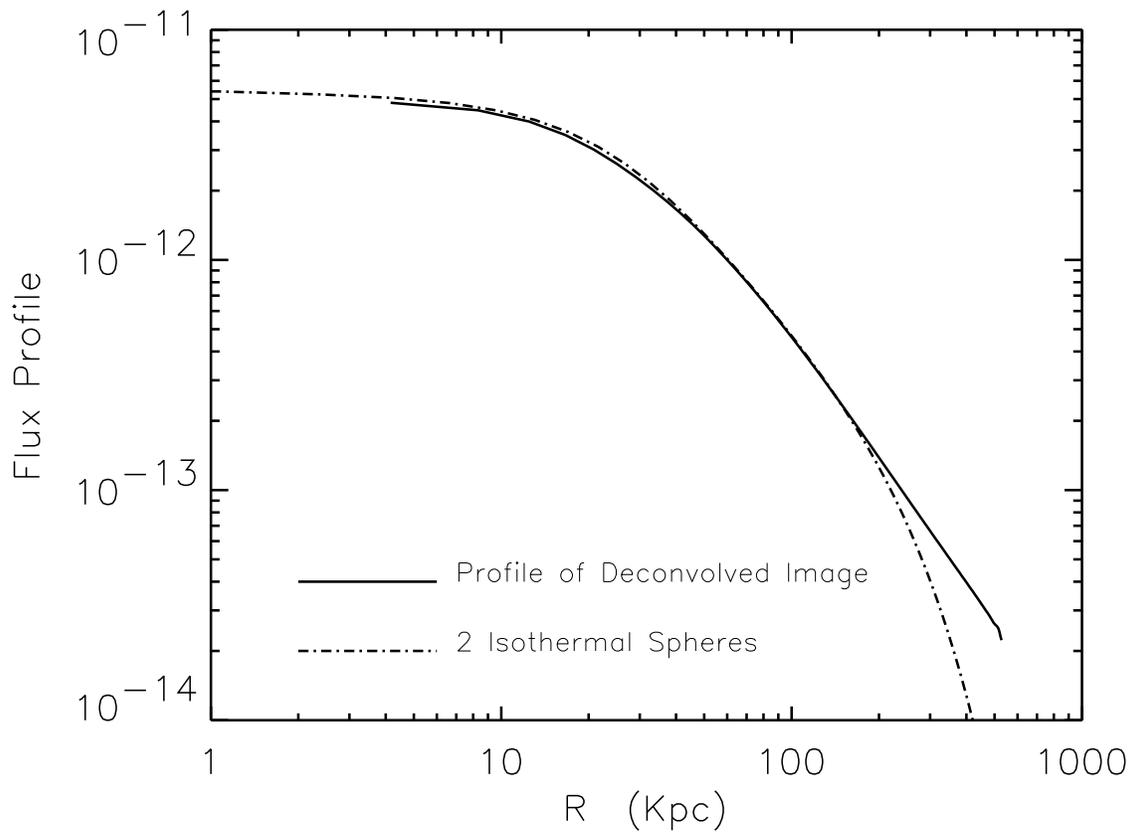}
\figcaption{X-ray Brightness profile.  2 concentric isothermal mass spheres
with $r_c=26.7$ h$_{50}^{-1}$ kpc and 1 h$_{50}^{-1}$ Mpc, central
density ratios 150:1 (dash-dotted line). The solid line is the
earlier $\beta$ model fit to deconvolved data.
}
\label{2isosflux}
\end{figure}

\clearpage
\begin{deluxetable}{rcrrl}
\footnotesize
\tablecaption{Photometric and spectroscopic data for the \ms04 field.
\label{tbl-1}}
\tablewidth{350pt}
\tablehead{
\colhead{Flag} & \colhead{R {\it mag/arcsec$^{2}$}} & \colhead{cz
{\it km/s}} &  \colhead{cz {\it err}} & \colhead{z}}
\startdata
*G1  &       & 57414 &   62  &  0.1915 \nl
*G2  & 19.34 & 58551 &  160  &  0.1953 \nl
*G3  & 18.52 & 59563 &   39  &  0.1987 \nl
*G4  & 20.87 & 58953 &  153  &  0.1966 \nl
*G5  & 19.28 & 58437 &   91  &  0.1949 \nl
*G6  &       & 57188 &   74  &  0.1908 \nl
*G7  &       & 59070 &  200  &  0.1970 \nl
*G8  & 18.41 & 59772 &   43  &  0.1994 \nl
*G9  & 18.72 & 56352 &   41  &  0.1880 \nl
*G10 &       & 59525 &   65  &  0.1985 \nl
*G11 &       & 60655 &   90  &  0.2023 \nl
 G12 & 19.96 & 54799 &  100  &  0.1828 \nl
*G13 & 17.70 & 56486 &   50  &  0.1884 \nl
*G14 & 19.58 & 59291 &   91  &  0.1978 \nl
*G15 &       & 60058 &  103  &  0.2003 \nl
 G16 & 18.32 & 54426 &   52  &  0.1815 \nl
*G17 & 18.36 & 59036 &   79  &  0.1969 \nl
*G18 &       & 58780 &   61  &  0.1961 \nl
*G19 &       & 59098 &   70  &  0.1971 \nl
*G20 &       & 60193 &  203  &  0.2008 \nl
*G21 &       & 59831 &   84  &  0.1996 \nl
 G22 & 20.25 & 54094 &  128  &  0.1804 \nl
*G23 & 20.12 & 60051 &  105  &  0.2003 \nl
*G24 &       & 57274 &   95  &  0.1910 \nl
*G25 &       & 59834 &  133  &  0.1996 \nl
*G26 &       & 57565 &   93  &  0.1920 \nl
 G27 &       & 54998 &  176  &  0.1834 \nl
*G28 &       & 59459 &   82  &  0.1983 \nl
*G29 & 20.82 & 58366 &   81  &  0.1947 \nl
*G30 & 18.21 & 57797 &   53  &  0.1928 \nl
*G31 &       & 59256 &  175  &  0.1977 \nl
*G32 & 20.28 & 60908 &   79  &  0.2032 \nl
*G33 & 17.36 & 59539 &  103  &  0.1986 \nl
*G34 &       & 57655 &   84  &  0.1923 \nl
*G35 & 19.25 & 59557 &   90  &  0.1987 \nl
*G36 &       & 59229 &  103  &  0.1976 \nl
*G37 & 18.21 & 59169 &   69  &  0.1974 \nl
  A1 &       & 159404 &  89  &  0.5317 \nl
 *C1 &       & 57532 &   290 &  0.1929 \nl
 *C2 &       & 59031 &   290 &  0.1969 \nl
 *C3 &       & 59960 &       &  0.20   \nl
 *C4 &       & 58221 &   290 &  0.1942 \nl
 *K1 & 20.26 & 58821 &    89 &  0.1962 \nl
 *K2 & 19.33 & 59510 &   119 &  0.1985 \nl
 *K3 & 19.38 & 59091 &   150 &  0.1971 \nl
  K4 & 17.26 & star  &       &  M      \nl
  K5 & 17.21 & 141476 & 180  &  0.4719 \nl
  K6 &       & 160303 &      &  0.5347 \nl
  K7 &       & 115993 &  90  &  0.3869 \nl
  K8 &       & 232915 &  30  &  0.7769 \nl
  K9 &       & 149690 & 600  &  0.4993 \nl
   1 & 20.21 & 78458 &  140  &  0.2617 \nl
   2 &       & 34581 &   81  &  0.1153 \nl
   3 &       & 28236 &   78  &  0.0942 \nl
   4 &       & 27374 &  149  &  0.0913 \nl
   5 &       & 51428 &  108  &  0.1715 \nl
   6 &       & 22960 &   83  &  0.0766 \nl
   7 & 21.90 & 42203 &  124  &  0.1408 \nl
   8 &       & 42255 &  116  &  0.1409 \nl
\enddata
\end{deluxetable}

\end{document}